# The Innovation Paradox: Concept Space Expansion with Diminishing Originality and the Promise of Creative AI

**Serhad Sarica**
Singapore University of Technology and Design
serhadsarica@gmail.com

**Jianxi Luo**
City University of Hong Kong
jianxi.luo@cityu.edu.hk

Version: 25 March 2024

**Abstract**

Innovation, typically spurred by reusing, recombining, and synthesizing existing concepts, is expected to result in an exponential growth of the concept space over time. However, our statistical analysis of TechNet, which is a comprehensive technology semantic network encompassing over four million concepts derived from patent texts, reveals a linear rather than exponential expansion of the overall technological concept space. Moreover, there is a notable decline in the originality of newly created concepts. These trends can be attributed to the constraints of human cognitive abilities to innovate beyond an ever-growing space of prior art, among other factors. Integrating creative artificial intelligence (CAI) into the innovation process holds the potential to overcome these limitations and alter the observed trends in the future.

## 1. Introduction

Innovation leads to new technological concepts and expands the cumulative space of technological concepts. Following the combinational view of creativity (Arthur 2007; Uzzi et al. 2013; He & Luo 2017; Han et al. 2018), new technological concepts serve as new building blocks for future inventors to recombine and synthesize into even newer ones. This process of new concepts empowering newer concept creation suggests a cycle of positive reinforcement and increasing returns of innovation (Arthur 1989). Therefore, new concept creation is expected to accelerate and result in exponential expansion of the concept space over time.

However, the accumulation of technological concepts created through innovation over time may increase the knowledge demands or burdens on future innovators (Jones 2009). To derive originality beyond an expanding space of prior art, later-coming innovators need to navigate, learn, master, synthesize, and benchmark against a wider space of prior art than before, engage more multidisciplinary teams, and cope with increasing complexity and uncertainty in the invention process (Luo & Wood 2017). As a result, achieving originality in design might become more difficult. Such negative reinforcement may slow down innovation over time.

Therefore, the cumulative expansion of the total technology space, resulting from innovation over time, may exert two opposing forces on future innovation, as depicted in Figure 1. On one hand, the expansion of the technology space due to innovation provides more technological concepts for future innovators to reuse, recombine, and synthesize into newer ones, which could accelerate innovation and further expand the technology space. On the other hand, this expansion increases knowledge prerequisites on future inventors and raises the bar for deriving design originality, consequently decelerating both innovation and the expansion of the technological concept space. We refer to this as the innovation paradox, which is the focal point of this paper.

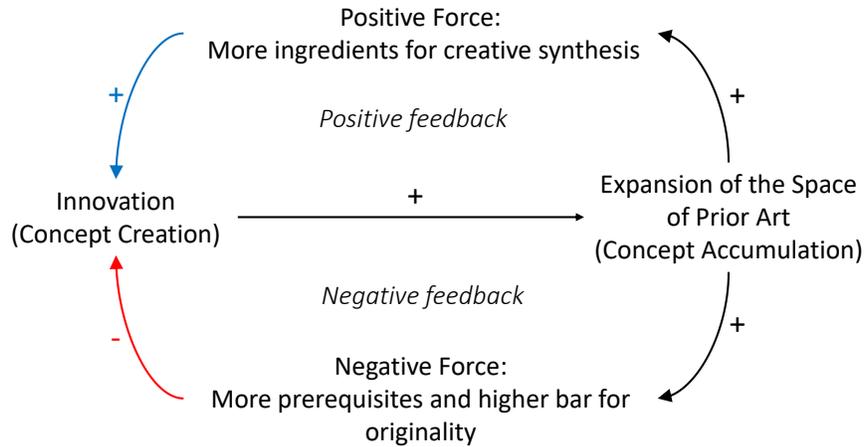

**Fig. 1.** The Innovation Paradox: Interplay of Positive and Negative Feedback in the Creation and Accumulation of Technological Concepts

In this study, we are particularly interested in the pattern and pace of growth of the technology space through innovation. More specifically, we ask - *is the total technology space expanding exponentially?* If the technology space growth is not exponential, it implies that innovation is decelerating, and various mechanisms are significantly limiting the combinatorial potential of the technology space. Such understanding could have implications for the future of innovation processes, informing the methods that innovators could employ, the strategies that firms might adopt, and the policies that governments should consider to sustain innovation.

Herein, we attempt to answer this question by statistically analyzing the scale and structural evolution of the technology semantic network, which we refer to as TechNet (Sarica et al. 2020), as a proxy for the total technological concept space. TechNet comprises over 4 million unique technical terms, including words and phrases, extracted from the texts of all granted utility patents in the complete United States Patent and Trademark Office (USPTO) database. These terms represent elemental technological concepts created throughout history, spanning all domains of technology. Specifically, we assess the originality of these once-new concepts at the time they first appeared in the cumulative technological concept space, according to their semantic similarity with prior concepts based on graph theory, and the new information content they add to the technological concept space based on information theory. Our results show a linear rather than exponential expansion of the overall technological concept space and a continual decrease in the originality of new concepts when created, indicating a possible deceleration of innovation.

In the following, we first review the relevant literature in Section 2, followed by an outline of our research methodology in Section 3. Our findings, accompanied by interpretations, are presented in Section 4. Section 5 delves into the potential of creative artificial intelligence in altering the observed trends. Subsequently, Section 6 discusses the limitations of our data and methodology. We draw conclusions in Section 7.

## 2. Related Work

### 2.1 The Deceleration of Innovation

Earlier studies have provided coarse-grained empirical evidence on possible declining trends in innovation. For instance, Huebner (2005) discovers a declining trend of breakthrough inventions over time based on the count of noticeable innovations in history. Jones (2009) observes an increase in inventors' age at their first invention, which suggests that more education and exploration time are required before actualizing originality in their ideas. Luo and Wood (2017) find a decrease in the number of patents per average inventor, indicating a decline in inventive productivity. Bloom et al. (2020) report declining research productivity based on firm- and industry-level census data in the semiconductor, crops, and health industries. Park et al. (2023) investigates 45 million papers and 3.9 million patents between 1945 and 2010 and how they change citation networks to prior work over time as a way to measure their disruptive impact. Their results show a declining trend of such impact for both papers and patents.

*2.2 Measures of Innovation*

These prior studies commonly investigate patent data but only employ measures that are extrinsic to innovations to detect potential trends in innovations. A technological innovation often appears and gets recognized as a new method, tool, product, system, service, or artifact. The newness or originality distinguishes it as an innovation from normal designs or artifacts, but not all elements of an innovation need to be original (Sternberg & Lubart 1999; Simonton 1999; Kaufman and Baer 2004; He & Luo 2017). In this study, we aim to look inside innovations and investigate the elemental concepts that constitute innovations and contribute to their originality. More specifically, we focus on measuring the originality of new concepts when they appear for the first time in the total technological concept space to investigate innovation trends over time. Note that the term "originality" is often used interchangeably with "novelty" in literature. Hereafter, we use "originality" for consistency.

Prior studies present various ways to assess originality in innovation. Creative design evaluation is mostly carried out by experts or other types of human subjects, with or without structured guidance or procedures (Sarkar & Chakrabarti 2007; Brown 2015; Ahmed & Fuge 2018; Sosa 2019; Hay et al. 2019). However, human experts are subject to their personal knowledge, experiences, opinions, and intuitions (Weisberg 2006; Oman et al. 2013). What appears original to one group of experts might not be so to others. Ideally, the originality of a technological concept should be assessed with reference to all previous concepts in history (Boden 1996). Experts may have incomplete knowledge of prior art, insufficient cognitive capacity to assess numerous new concepts, ideas, or innovations—e.g., at the magnitude of millions—and are unlikely to provide statistical significance in their assessments.

To cope with the limitations of human evaluation, data-driven and computational approaches have been developed to evaluate the originality of designs, inventions, creative design ideas, and so on (Luo 2023a). Many of these approaches are based on patent databases. Patents contain rich design or technological information and have been widely used as proxies for technological innovations in empirical studies. Patent data have been extensively employed as digital design repositories to develop engineering design theories and data-driven design support methods and tools (Jiang et al. 2022), although patent data sources are subject to a complex institutional system and many non-technical factors, such as examination processes, policy changes, and industry differences.

Patent databases host millions of patent documents in all technological domains and thus can support statistically significant, rigorous, and systematic data-driven evaluations of inventiveness or originality. Fleming & Sorenson (2001), Kim et al. (2016), and He & Luo (2017) statistically analyze the rareness of historical co-occurrences of co-classes of patents (or the classes of patent references) to measure the originality of patented inventions from a recombination perspective. Uzzi et al. (2013) assess the rareness of the combinations of references of millions of academic papers to provide statistically significant indicators of the originality of published research.

*2.3 Natural Language Processing*

More recent studies have leveraged large pre-trained lexical databases and natural language processing (NLP) techniques to automatically assess the originality of newly generated design concepts in natural language descriptions (Siddharth et al 2022a). Siddharth et al. (2020) propose a method to evaluate the novelty of a design solution, based on its distance to a reference product database. This distance is calculated as text similarity using the SAPPhIRE model. Camburn et al. (2020) evaluate the novelty of a large quantity of crowdsourced design ideas according to the semantic distance among the terms in the design idea description texts, with semantic distance derived from Freebase. Han et al. (2020) employ ConceptNet to assess the novelty of new design ideas based on the semantic distance between elemental concepts. Gerken and Moehrle (2012) used Subject-Action-Object triplets to measure the semantic similarity between patents to indicate the novelty of patents concerning others. Olson et al. (2021) similarly used a measure of semantic distance among words that a person generates as an indicator of thought space divergence and creativity.

To calculate the semantic distance (or similarity) between words or phrases, a comprehensive knowledge base is necessary. WordNet and ConceptNet have been the most used knowledge bases for this purpose (Linsey et al. 2012; Kan & Gero 2018; Georgiev & Georgiev 2018; Goucher-Lambert & Cagan 2019; Han et al. 2022). TechNet is a relatively newer knowledge base, trained on engineering design data, such as patent texts, and is more suitable than ConceptNet, WordNet, and other common-sense knowledge bases for assessing semantic distance or similarity between technological concepts (Sarica et al. 2020; 2023).

On the other hand, the originality of terms used in a patent or an academic paper can serve as an indicator of the originality of the described design or research finding. New terms, compared to prior terms in the literature or technological knowledge base, represent original concepts that may challenge the status quo and shape the future directions of the corresponding field (Kuhn 1970; Wray 2011). For instance, Park et al. (2023) discover a trend of declining new word-pair occurrences in patent titles, indicating a decrease in the diversity of word usage and a decline in combinatorial novelty. Originality is a matter of degree.

In this study, we employ TechNet, a large semantic network of technical concepts, as the semantic knowledge base for assessing the degree of originality of technological concepts created over time. The originality assessment centers on the semantic distance (the opposite of semantic similarity) between new terms in new patents and prior terms in earlier patents, using several metrics based on graph theory and information theory. Consequently, our analysis focuses on concepts in patentable technological inventions from the past four decades. In the following sections, we introduce TechNet, the patent data source, and our metrics in detail.

## 3. Measuring Concept Creation and Originality in Technology Semantic Network

### 3.1 Technology Semantic Network (TechNet)

The technology semantic network utilized for this research, termed TechNet, has been pre-trained and published in our prior work (Sarica et al. 2020). It is made publicly available on the web (https://www.tech-net.org) for external researchers, allowing access for broader research and applications. TechNet is a semantic network comprising of 4,038,924 technology concepts (words and phrases up to 4-grams) and their pairwise semantic distances. To ensure comprehensive coverage of technology concepts across all technology and engineering domains, the source data for constructing TechNet included all 5,771,030 utility patents granted between 1976 and October 2017 from the USPTO database.

Due to incomplete data coverage for the year 2017, we chose not to include data from that year. Furthermore, the digital patent database only became available starting in 1976. Since concepts created before 1976 could be reused in innovation and appear in patents after 1976, the "new" concepts in the initial years of the database may not be genuinely new, but only new to the database. Our analysis reveals that concepts appearing in patents between 1976 and the end of 1980 constitute over 90% of the total concepts used throughout the entire period from 1976 to 2016, indicating comprehensive coverage of baseline patent knowledge. Consequently, we established an initial semantic network, including concepts that appeared in patents between 1976 and the end of 1980 as the baseline. Our trend analysis commences in 1981.

In constructing TechNet, we began by extracting words and phrases from the raw texts of patent titles and abstracts. These extracts represent meaningful technological concepts (e.g., functions, components, structures, materials, configurations, working principles) and were processed using natural language processing techniques for phrasing, denoising, lemmatizing, among others. The original text database contains 26,756,162 sentences and approximately 699 million words. Utilizing this data, several word embedding models were trained on the preprocessed sentences to derive the embedding vectors of individual concepts, thereby forming a unified embedding vector space representing the total technological concept space. A technological semantic network was then constructed by connecting these technological concepts based on the cosine similarity of their embedding vectors, i.e., semantic similarity or an inverse indicator of semantic distance. By using the total database to train the embedding space of concepts, we anticipate that the embedding vector similarity will reflect the most intrinsic technical relations between the technical concepts represented by the terms.

Sarica et al. (2020) performed a benchmark comparison with other large semantic networks and knowledge databases (e.g., WordNet and ConceptNet, most of which were trained on common-sense databases) and found that the word2vec embedding model (Mikolov et al. 2013) yielded the best-performing technology semantic network for concept retrieval and inference tasks within the specific context of technology and engineering. Consequently, the technology semantic network (TechNet) used in this study is based on *word2vec*.[1]

Figure 2 shows an example subgraph consisting of 30 concepts sampled from the technology semantic network cumulative to the year of 1990. With an interest in new concept creation and the evolution of concept space over time, we created the longitudinal technology concept network cumulative to each year from 1980 to 2016 and identified the concepts that appeared for the first time and were "original" in each respective year. The associations of subsets of concepts in the yearly networks are based on their pairwise semantic similarity (i.e., embedding vector similarity) and are derived from the total embedding space trained on the complete patent database from 1976 to 2016. Therefore, although the concepts included in yearly networks change, their pairwise semantic similarity is universal. In the next section, we introduce the graph-theoretic and information-theoretic metrics in our analysis based on TechNet.

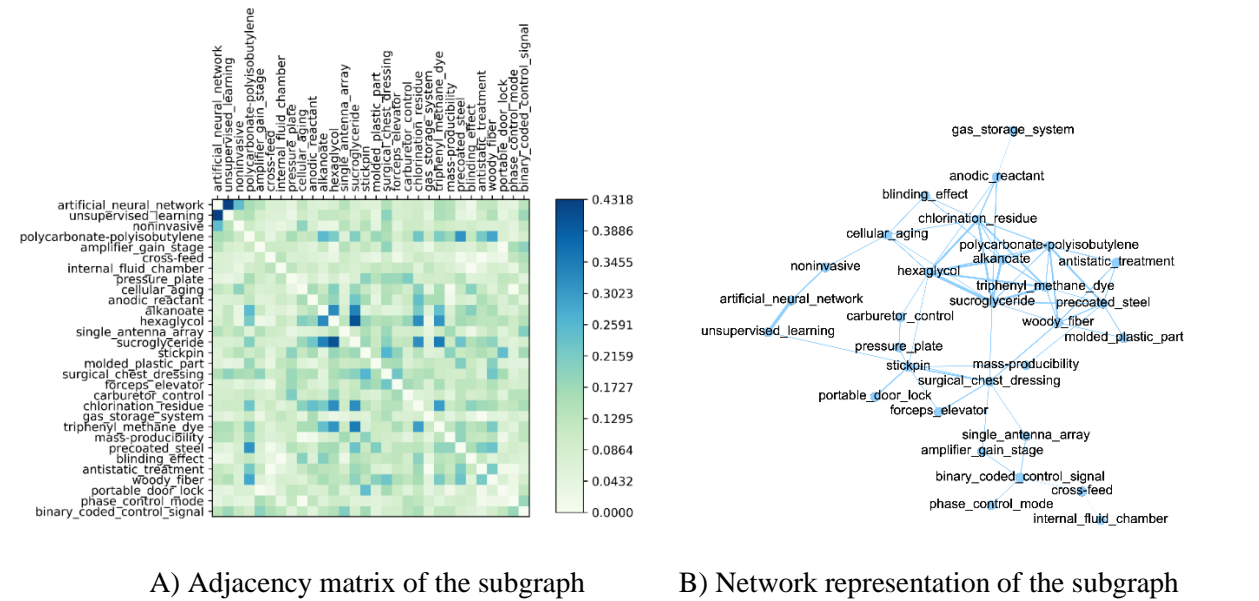

A) Adjacency matrix of the subgraph        B) Network representation of the subgraph

---

**Fig. 2.** An example subgraph of 30 concepts sampled from the total technology concept network cumulative to 1990. (A) is the adjacency matrix representation of the subgraph where the value of each cell is the semantic similarity of the corresponding tuple. (B) is a filtered network representation of the subgraph. In the total concept network cumulative to 1990, the share of the new concepts in cumulative total concepts is 5.4%. Preserving this ratio, the sample subgraph has 2 new concepts and 28 prior concepts. The concepts "artificial neural network" and "unsupervised learning" appeared for the first time in 1990, whereas the other 28 concepts had occurred in previous years.

*3.2 Graph and Information Theoretic Metrics*

a) The network of concepts and originality of new concepts

For a concept network $G = (V, E)$, let two concepts be $v_i$ and $v_j \in V$. The semantic similarity between them is denoted by $w_{ij}$. Then, the mean semantic similarity of concepts in the network (representing a concept space) is calculated as

$$w_G = \frac{2}{N(N-1)} \sum_{i,j,i \neq j} w_{ij} \qquad (1)$$

where $N$ is the number of concepts in the network.

$w_G$ is an inverse indicator of the divergence of the concept space. Recently, Olson et al. (2021) similarly used a measure of the average semantic distance among words that a person generates as the indicator of thought space divergence and creativity. This measure will further allow us to detect whether the cumulative technological concept space has been diverging or converging over time, as new concepts continually enter the space each year.

We further assess the mean semantic similarity $w_N$ between the new and prior concepts as

$$w_N = \frac{1}{|U||V|} \sum_{\forall i \in U, \forall j \in V} w_{ij} \qquad (2)$$

where $U$ is the set of concepts that appeared prior to the corresponding year and $V$ is the set of new concepts that appeared in the corresponding year. The calculation considers only the edges between the concepts in sets $U$ and $V$.

$w_N$ is an inverse indicator of the originality of the new concepts that appeared for the first time in a year. It follows the spirit of a few recent studies that similarly used measures of semantic distance between words or terms as novelty indicator of design ideas (Camburn et al. 2020; Goucher-Lambert & Cagan 2019). This metric will further allow us to detect the longitudinal change in the originality of the new concepts appearing for the first time each year.

Due to the large size of the total TechNet, for each year, we randomly sample 100 subgraphs of the total concept network cumulative to each year for calculating the graph theoretic metrics. Each subgraph contains 1,000 (or 500, 2000, 5000 in the robustness tests) randomly sampled concepts from the total network cumulative to each year. In each random subgraph, the share of new concepts is preserved to be the same as the share in the total network cumulative to that year.

b) Concept information content

We also measure the additional new information content that is brought by a new concept to the cumulative space of technological concepts. This can imply the amount of learning required to remove uncertainty over the meaning of the new concept. Assuming that all prior concepts have been known by the collective intelligence of human innovators, the information content of a new concept can be approximated as the sum of the information content of the most similar prior concept to the new concept and the additional information content that the new concept brings in. This can be expressed as,

$$IC(x_{new}) = IC(p) + \Delta IC(x_{new}|p) \qquad (3)$$

where $IC(x_{new})$ is the information content of the newly introduced concept $x_{new}$, $IC(p)$ is the information content of the prior concept $p$ most similar to $x_{new}$, and $\Delta IC(x_{new}|p)$ is the additional information content that $x_{new}$ brings to collective concept space given the most similar prior concept $p$.

According to Shannon's information entropy, the information content of an event $x$ is,

$$IC(x) = -\log P(x) \qquad (4)$$

It states that if an event $x$ has a lower probability of occurrence, i.e., if $P(x)$ is lower, its information content is higher. Thus, an unexpected event carries more information than a highly expected event.

By inserting equation (4) into equation (3), we obtain $\Delta IC(x_{new}|p)$ as:

$$\Delta IC(x_{new}|p) = \log \frac{P(p)}{P(x_{new})} \qquad (5)$$

Since it is assumed that prior concepts are collectively known, $P(p)$ can be approximated as 1 (i.e., the probability that someone is an expert on prior concept $p$ is 1). On the other hand, $P(x_{new})$ can be approximated as the cosine similarity between the new concept and its most similar prior concept. In other words, the probability of inferring the meaning of the new concept is approximated by its maximum similarity to prior concepts.

Equation (5) with log base 2 is used to calculate each new concept's expected additional information content when it appeared for the first time. On this basis, we measure the mean additional information content that a sample of 1,000 (or 500, 2,000 and 5,000 in the robustness tests) randomly sampled new concepts bring to the technology semantic network each year. Then we detect the changes of the mean additional information content of new concepts each year.

Our data and codes are available at https://github.com/SerhadS/techspace-evolution.

## 4. Findings and Interpretations

*4.1 Findings*

We now report the longitudinal changes in the macro- and micro-structures of the technological concept network over the past four decades, focusing on the originality and information content addition of new concepts that appeared for the first time in the technological concept space cumulative to each year.

First, we observe that the number of accumulated concepts within the total technological concept space has exhibited linear growth, signifying a relatively steady annual creation of new concepts. The annual growth rate of new concepts has consistently decreased over the past four decades (Figure 3).[2] This rate is calculated by dividing the number of new concepts in each year by the cumulative number of concepts in the total space up to that year. Had the growth rate been constant or increasing, the cumulative curve would have taken an exponential shape. Moreover, the ratio of new concepts to all unique concepts, assessed in rolling 5-year windows, has also seen a decline (refer to Supplementary Materials). Concurrently, there has been a reduction in the average number of new concepts per patent, decreasing from 2.15 to 0.54, while the total number of concepts per patent has remained relatively stable. In summary, the expansion of the total technological concept space is linear, rather than exponential.

---

[2] The pattern here should not be confused with the one based on Heap's law. Heap's law describes the relationship between the number of distinct terms from a vocabulary in a text and the length of the text. Our case addresses the relationship between the number of new terms entering a vocabulary and the expanding size of the vocabulary. The length of individual patent abstract texts has been rather stable.

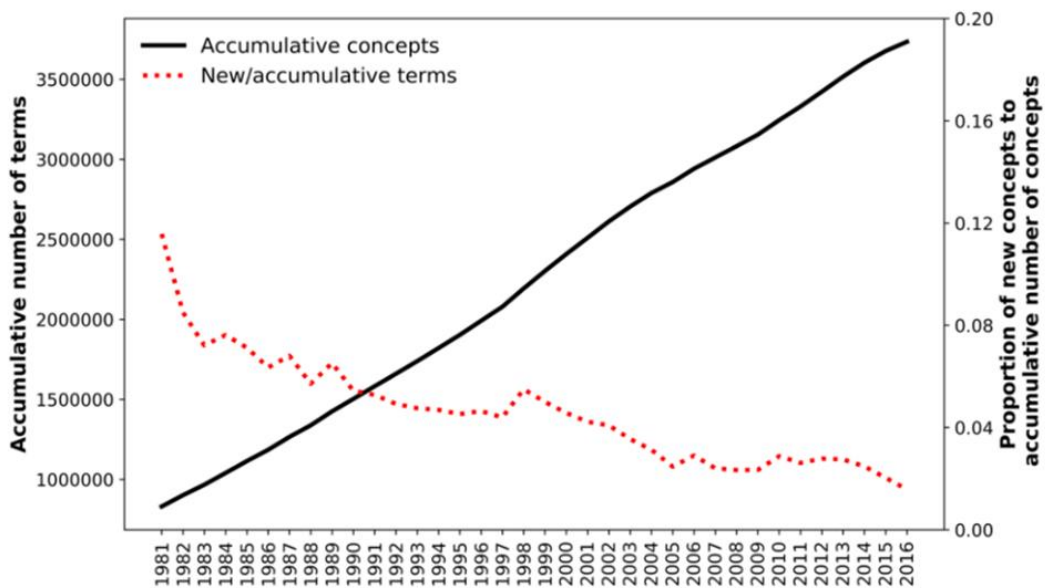

**Fig. 3.** The total number of concepts and the proportion of new concepts to the total number of concepts in the network, accumulated up to a given year.

Second, as shown in Figure 4, from 1981 to 2016, the mean semantic similarity of all concepts increased by 23%. This suggests that the concepts in the technological concept space are converging over time. We conducted Kolmogorov–Smirnov tests, and the results show significant increases across years and periods of 5 years (see Supplementary Materials). Furthermore, the mean semantic similarity between new and prior concepts increased by 31%.[3] This suggests that concepts created each year (i.e., appearing for the first time in the network cumulative to that year) are becoming more similar to prior concepts, implying that the originality of new concepts is diminishing over time. To test the robustness of this pattern, we also experimented with samples containing 500, 2,000, and 5,000 concepts to calculate mean semantic similarity, and the results are reported in Figure 5. The trend remains consistent.

---

[3] As a relevant test, we randomly chose 10,000 patents every year and calculated the pairwise semantic similarity between unique terms in patent titles and abstracts. We find a slight increase in the mean semantic similarity among the concepts within individual patents from 0.27 to 0.31, with a rather stable number of total concepts per patent. See Supplementary Materials.

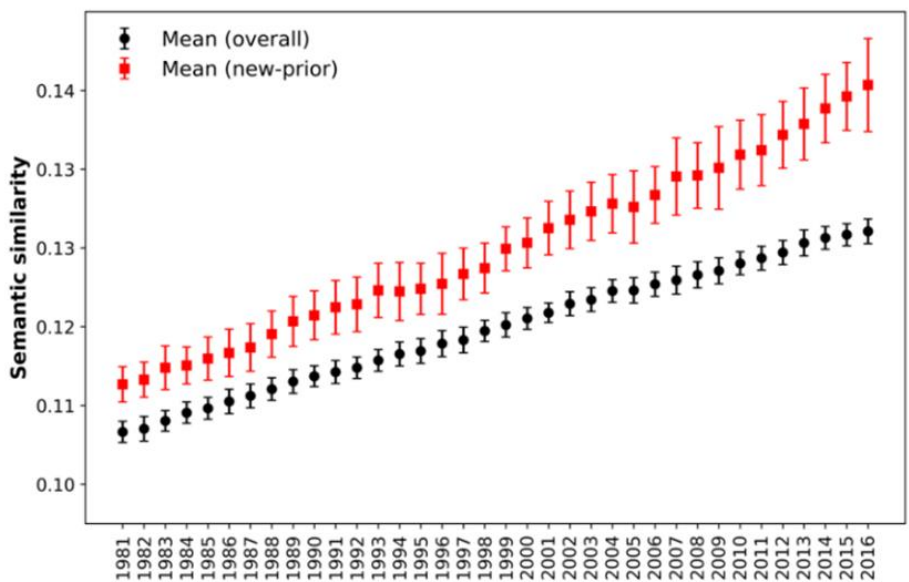

**Fig. 4.** The mean semantic similarity of all concepts and the mean semantic similarity between new and prior concepts in the network accumulated up to a given year. Due to the size of the technology concept network, for computational efficiency, we sampled 100 subgraphs, each comprising 1,000 randomly selected concepts, from the total network accumulated up to each year, and calculated the means and standard deviations of the mean semantic similarity for the 100 subgraphs.

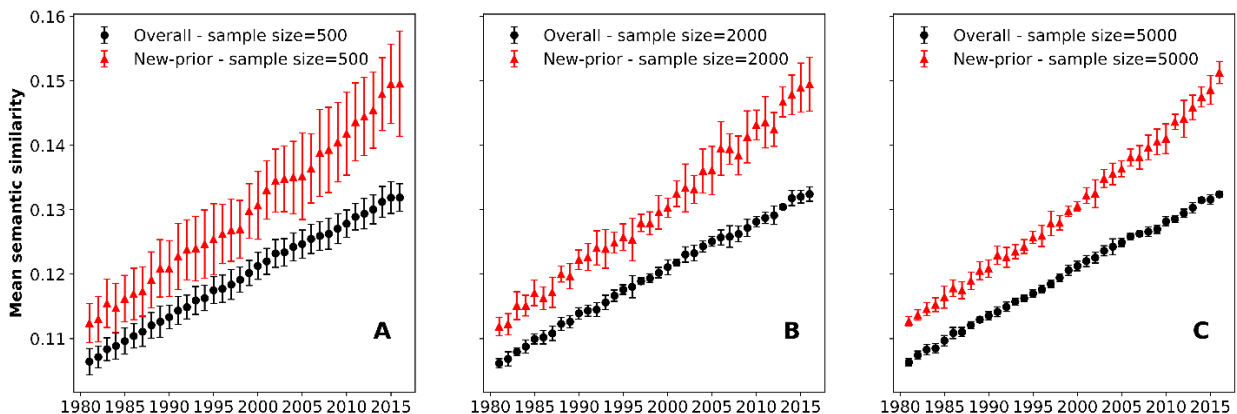

**Fig. 5.** Robustness tests for mean semantic similarity measurement. The mean (node) and standard deviation (error bar) of semantic similarities of the concepts in 100 randomly sampled subgraphs, each consisting of (A) 500 concepts, (B) 2,000 concepts, (C) 5,000 concepts in each year. The differences between sub-plots suggest higher variance for smaller subgraph sizes and lower variance for larger subgraphs, as expected.

Third, Figure 6 shows a continuous 21% decrease in the mean additional information content that an average new concept contributes to the prior total concept space from 1981 to 2016. We conducted Kolmogorov-Smirnov tests, and the results demonstrate that the decreases across years and periods of 5 years are significant (see Supplementary Materials). To test the robustness of this trend, we experimented with samples containing 500, 2,000 and 5,000 concepts to calculate mean additional information content, and the results are displayed in Figure 7. This consistent trend suggests that the newly created concepts are contributing a diminishing amount of new information to the existing knowledge base. For example, when the term "deep learning" first appeared in TechNet, it was highly associated with many previously existing concepts, such as "neural network", "machine learning", and "regression", thus adding little new information to the prior concept space. The same applies to "blockchain", "Web 3", "metaverse", and many other emerging terms, which can be considered as "rehashed concepts" with low originality.

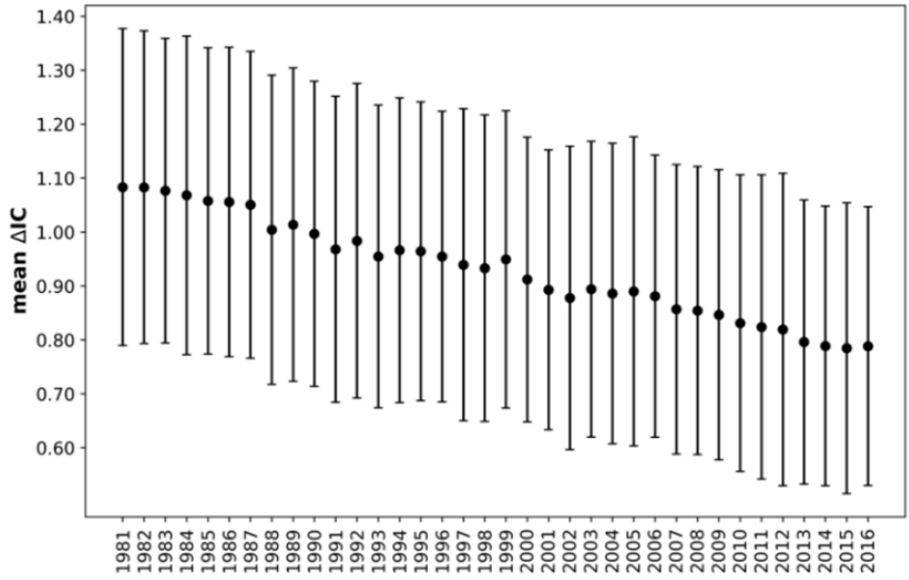

**Fig. 6.** The mean additional information content contributed by 1,000 randomly selected new concepts to the technology concept network. The means and standard deviations are denoted by the nodes and error bars, respectively.

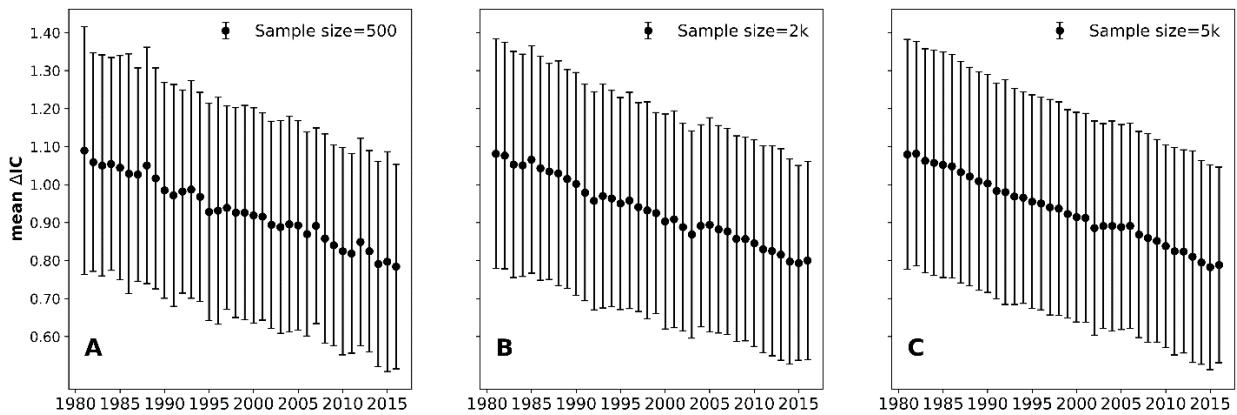

**Fig. 7.** Robustness tests for mean additional information content measurement. Longitudinal change in mean (node) and standard deviation (error bar) additional information content brought by new concepts in samples of (A) 500 concepts, (B) 2,000 concepts, (C) 5,000 concepts in each year. Although the sub-plots are similar, smaller samples exhibit slight fluctuations, which diminish in larger ones.

Similar or related trends have been documented in several prior studies, albeit with more coarse-grained empirical evidence at the level of discrete breakthrough inventions (Huebner 2005), patented inventions (Luo & Wood 2017), or regarding the behaviors and productivity of individuals (Jones 2009), firms, and industries (Bloom et al. 2020). Our analysis examines the elemental concepts that constitute inventions and their relations that form the total technology space, thereby providing both finer-grained and more macro empirical evidence on the possible trends of innovation. Notably, our work is enabled by the latest natural language processing technologies and, specifically, the large pre-trained technology semantic network, TechNet, which has only recently become publicly available.

### 4.2 Interpretations

The linear expansion of the total technological concept space, coupled with the shrinking originality of new concepts over time, implies diminishing returns on the creation of new concepts. This trend may stem from the dominant negative feedback illustrated in Figure 1, which outweighs the positive feedback.

The cumulative expansion of the total technological concept space as a result of continual innovation implies more knowledge prerequisites for future innovators (Jones 2009; Callaghan 2021) and more prior concepts to benchmark against to derive originality in future innovation. To invent new technologies and create original concepts against an expanding space of prior art, future innovators must learn, master, utilize, and synthesize an ever-increasing number of prior concepts and knowledge than ever before. Many new concepts are "rehashed concepts" rather than truly original ones and might confuse the learning of young students and future innovators. Additionally, the increasing complexities in new technologies, design processes, and organizations, as evidenced and analyzed in many prior studies (de Weck et al. 2011; Luo & Wood 2017), may also introduce further challenges to future innovation.

On the other hand, the accumulation and expansion of the total technology space may offer more knowledge ingredients for potential reuse, recombination, and synthesis into new concepts. This is the positive feedback depicted in Figure 1. However, such creative recombination potentials are conditioned on the cognitive capabilities of humans to fully utilize, learn, and synthesize the growing space of prior concepts. The linear expansion of the technological concept space and diminishing originality of new concepts characterize the era to date when innovation primarily relied on biologically limited human intelligence. Moving forward, artificial intelligence (AI) may help cope with the growing knowledge burden and "complexity" challenges to human intelligence for innovation.

## 5. The Promise of Creative AI

Here, we introduce "Creative Artificial Intelligence" (CAI) as a type of AI with the potential to counteract the negative forces impacting innovation. CAI extends beyond traditional machine learning, which is primarily focused on pattern recognition, and transcends the realm of computer-aided design (CAD) and human concept generation capabilities (Nagai et al. 2009; English et al. 2010; Luo et al. 2018; He et al. 2019; Luo et al. 2021; Sarica et al. 2021). Creative AI distinguishes itself from Generative AI by necessitating that its outputs be both original and useful to qualify as genuinely creative. CAI needs to integrate three interlinked, knowledge-based capabilities essential for creative tasks: machine learning, machine creation, and machine evaluation (see Figure 8).

Machine learning absorbs prior knowledge and concepts efficiently, addressing the increasing knowledge burdens and complexity. Machine creation, through the automated recombination of existing concepts, fosters the generation of diverse and novel concepts, potentially counterbalancing the decline in originality. Machine evaluation swiftly compares new concepts against the vast repository of prior art, facilitating the identification of truly novel innovations (Haefner et al., 2021; Hutchinson, 2021; Luo, 2023a). These interconnected capabilities are vital for overcoming the challenges associated with the dwindling originality in innovation.

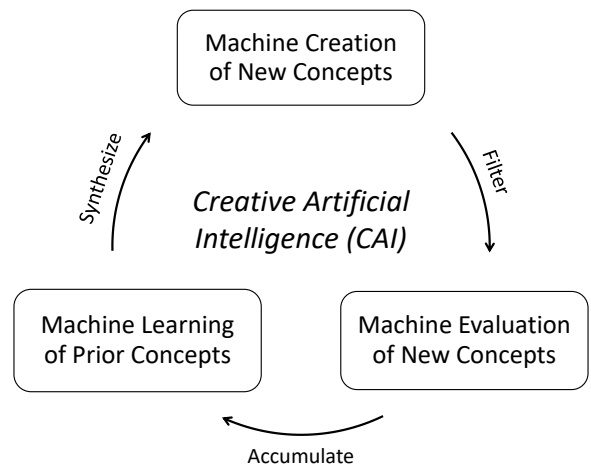

**Fig. 8.** The fundamental constituents of Creative Artificial Intelligence (CAI)

Recent breakthroughs in Generative Pretrained Transformers (GPTs) and Large Language Models (LLMs) have showcased extraordinary abilities in content generation from a broad knowledge base (OpenAI 2023). Although such capabilities do not inherently guarantee the originality of the generated content, they can be harnessed and refined into CAI through fine-tuning and creativity-oriented model controls. Our research at the Data-Driven Innovation Lab has leveraged design creativity theories to adapt or control various GPTs with curated design datasets, enabling them to achieve a higher level of artificial creativity that surpasses human ingenuity and generates highly original technological concepts for complex problems (Zhu & Luo 2023; Zhu et al. 2023). This underscores the promise of CAI to enhance creativity in the innovation process.

Furthermore, empirical studies have documented the exponential growth in the functional performance of key technology categories, notably in information and energy processing (Kurzweil 2005; Koh & Magee 2006; 2008). Such acceleration implies increasing returns on technological enhancements. These studies, however, concentrate on functional performance improvements rather than new concept creation and originality that drive innovation, the very elements our research focuses on.

The observed dichotomy indicates that while functional improvements have been accelerating, the origination of novel technological concepts has been decelerating. The past four decades suggest that technological progression has been predominantly incremental, prioritizing refinement over groundbreaking conceptual innovation. Nevertheless, original, pioneering innovations could still significantly influence long-term technological performance. On the other hand, the swift advancements in data computing, storage, and transmission technologies (Singh et al. 2021), identified as peaks in the technology fitness landscape (Jiang & Luo 2022), have propelled the development of GPTs and LLMs and may potentially fuel the CAI evolution.

Incorporating CAI into the innovation process holds promise for transforming the course of concept creation and originality, challenging the historical reliance on human intellect. As CAI assumes an increasingly prominent role in designing future technologies, we must ponder the future roles of human designers. The interplay between CAI and human intelligence in future design processes presents an urgent research inquiry (Song et al. 2022; Song et al. 2024). The co-design process involving humans and CAI must be thoughtfully crafted to safeguard fundamental human values, in both the innovation outcomes and the process itself (Luo 2023b). Nonetheless, it is crucial to recognize that societal, economic, geopolitical, and demographic factors may also influence the trends and future scenarios.

## 6. Limitations and Cautions

It is essential to exercise caution when interpreting and drawing inferences from the observed trends for several reasons related to data and methodology. First, our empirical basis is limited to patentable technological inventions during a specific period from 1976 to 2016 in a specific database from USPTO. Future research should examine non-patentable inventions and cover other time periods to test if the patterns we discovered hold true. While the U.S. patents provide the best proxy for measuring global invention originality to date, this situation might change in the future and examining the originality of patents from other nations might yield additional insights (Santacreu & Zhu 2018).

Second, the patent institutional system may introduce confounding factors that affect the interpretation of the trends we have observed over the past four decades. The propensity to patent varies over industries, organizations, and time periods. For example, many software inventions are not patented. Patent quality is also related to the examination process. Legal and institutional changes (e.g., the Bayh-Dole Act, the creation of the Patent Trial and Appeal Board in 1982) may impact the patent database. Future research should consider systematic statistical techniques to control for such confounding factors when testing specific interpretations.

Furthermore, our analysis is based on the only publicly available technology semantic knowledge base, TechNet, and several new metrics that we proposed. Future research may explore other and emerging semantic knowledge bases, such as the Engineering Knowledge Graph or EKG (Siddharth et al. 2022b), as well as alternative metrics to approximate the technological concept space and measure concept originality. Future research may also compare concept creation and originality patterns across different technological fields.

In sum, addressing these data and methodological limitations in future research may derive new insights and interpretations, and further support, nuance, or challenge our findings.

## 7. Conclusion

In this study, we analyzed the structure and evolution of the technology semantic network based on graph and information theories. Our results suggest diminishing originality of new concepts during linear expansion of the total concept space over the past four decades. We considered the negative and positive feedback (depicted in Figure 1) from prior innovation on future innovation to interpret these trends. Innovation over time results in the cumulative expansion of the technological concept space, which raises the bar for deriving originality in future innovation and increases knowledge burdens on future innovators, despite providing more knowledge ingredients for potential recombination and synthesis into new concepts. Developing and deploying creative artificial intelligence (CAI) in the innovation process might potentially strengthen the positive feedback and mitigate the negative feedback from prior innovation, thus altering the observed trends. The paper calls for more research on CAI and its impact on future innovation.

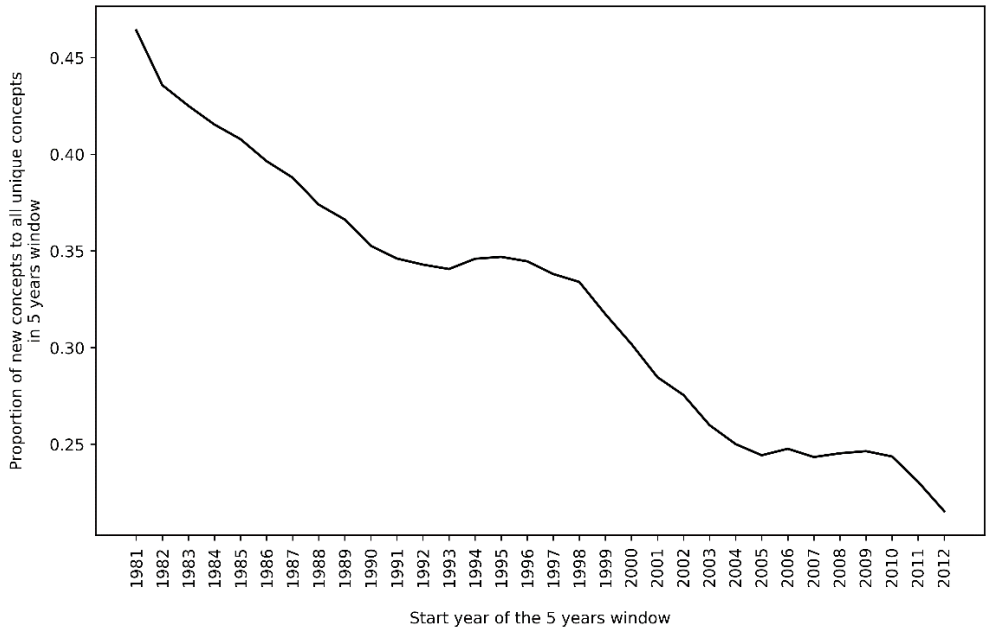

**Fig. S1.** The proportion of new concepts to all unique concepts in moving 5 years.

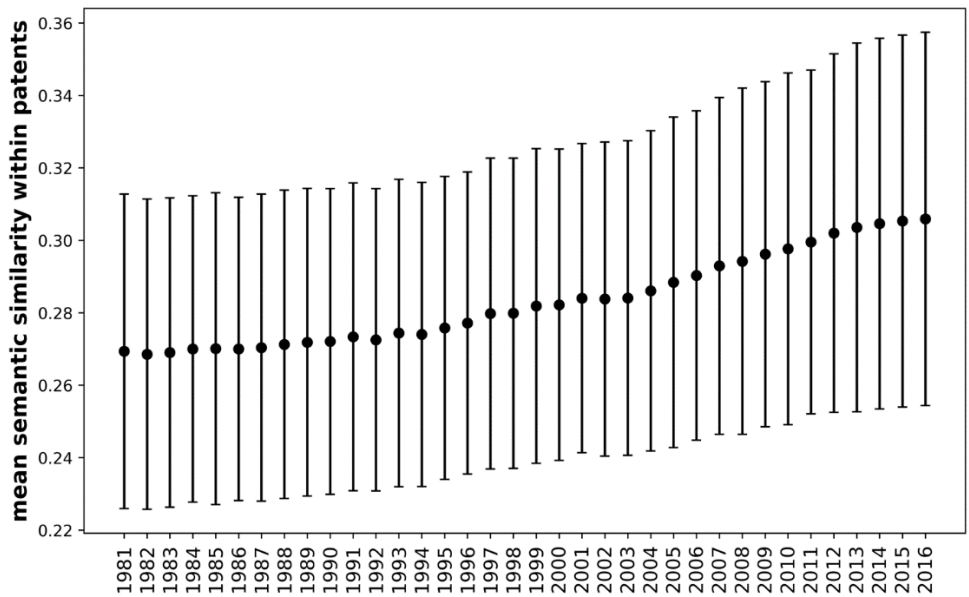

**Fig. S2.** Mean semantic similarity within 10,000 randomly chosen patents. We randomly chose 10,000 patents every year, calculated the pairwise semantic similarity between the unique terms within each patent title and abstracts, and took the mean of the distribution for each patent.

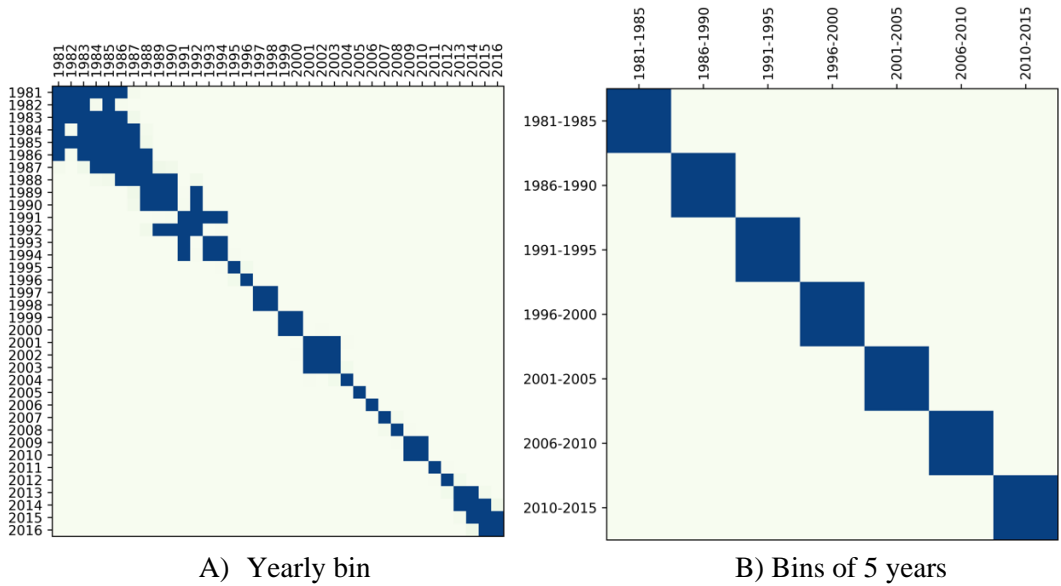

**Fig. S3.** Kolmogorow-Smirnov tests of the null hypothesis that two samples are drawn from the same distribution for mean semantic similarity distributions over time. The dark cells indicate that the distributions are not significantly different so that we cannot reject the null hypothesis. Plot A) shows that the difference is significant for years that have long periods of time in between, although several pairs of consecutive years seem to have similar distributions. Plot B) shows that the differences are significant across 5-year periods.

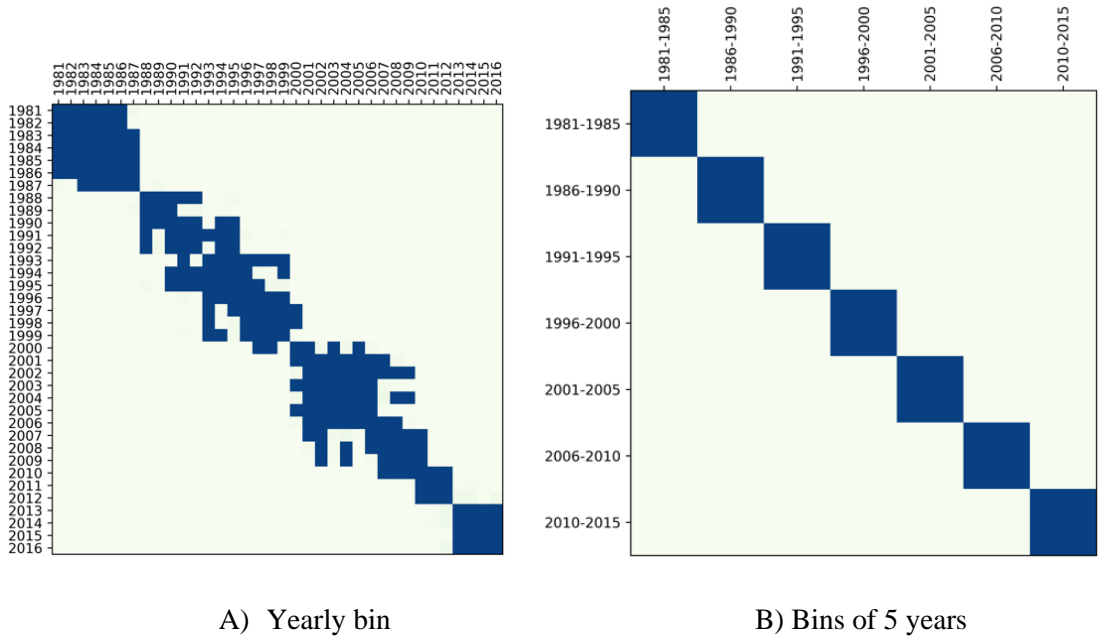

**Fig. S4.** Kolmogorow-Smirnov tests of the null hypothesis that two samples are drawn from the same distribution for delta information content distributions over time. The dark cells indicate that the distributions are not significantly different so that we cannot reject the null hypothesis. Plot A) shows that the difference is significant for years that have long periods of time in between, although several pairs of consecutive years seem to have similar distributions. Plot B) shows that the differences are significant across 5-year periods.